\documentclass[11pt,fleqn,twoside]{article}
\usepackage{amsfonts,amssymb,latexsym}
\makeatletter
\newcommand{\prava}[1]{\small\it
\begin{flushleft}
Copyright \copyright \ 1999 by  #1
\end{flushleft}}

\newcommand{\name}[1]{\begin{flushleft}
                       \LARGE \bf #1
                       \end{flushleft}\vspace{-3mm}}

\newcommand{\Author}[1]{\begin{flushleft}
                       \it #1 \end{flushleft}}

\newcommand{\Adress}[1]{\begin{flushleft}
                       \it #1 \end{flushleft}}

\newcommand{\Date}[1]{\begin{flushleft}
                      \small  \it #1 \end{flushleft}}

\newcommand{\ehkol}{Author \ name}
\newcommand{\ohkol}{Article \ name}
\renewcommand{\@evenhead}{
\hspace*{-3pt}\raisebox{-15pt}[\headheight][0pt]{\vbox{\hbox to \textwidth 
{\thepage \hfil \ehkol}\vskip4pt \hrule}}}
\renewcommand{\@oddhead}{
\hspace*{-3pt}\raisebox{-15pt}[\headheight][0pt]{\vbox{\hbox to \textwidth 
{\ohkol \hfil \thepage}\vskip4pt\hrule}}}
\renewcommand{\@evenfoot}{}
\renewcommand{\@oddfoot}{}

\newcommand{\be}{\begin{equation}}
\newcommand{\ee}{\end{equation}}
\newcommand{\ba}{\hspace*{-5pt}\begin{array}}
\newcommand{\ea}{\end{array}}

\makeatother

\def\textgoth#1{$\mathfrak{#1}$}
 
\newtheorem{proposition}{Proposition}[section] 
\newtheorem{theorem}{Theorem}[section] 
 
\begin{document} 
\thispagestyle{empty}

\renewcommand{\ehkol}{R. Martini and P.K.H. Gragert}
\renewcommand{\ohkol}{Solutions of WDVV Equations in Seiberg-Witten
Theory}

\begin{flushleft}
\footnotesize \sf
Journal of Nonlinear Mathematical Physics \qquad 1999, V.6, N~1,
\pageref{martini-fp}--\pageref{martini-lp}.
\hfill {\sc Letter}
\end{flushleft}

\vspace{-5mm}

\renewcommand{\footnoterule}{}
{\renewcommand{\thefootnote}{} 
\footnote{\prava{R. Martini and P.K.H. Gragert}}} 

\name{Solutions of WDVV Equations in Seiberg-Witten Theory from
Root Systems} \label{martini-fp}
 
\Author{R. MARTINI and P.K.H. GRAGERT}
 
\Adress{Faculty of Mathematical Sciences, University of Twente,\\
P.O. Box 217, 7500 AE Enschede, The Netherlands}
 
\Date{Received October 1, 1998;  Accepted November 14, 1998}
 
\begin{abstract} 
\noindent
We present a complete proof that solutions of the WDVV equations in
Seiberg-Witten theory may be constructed from root systems. A
generalization to weight systems is proposed.

\end{abstract}

{\advance\topsep-2pt 
\section{Introduction} 
Recently in $N=2$ four-dimensional supersymmetric Yang-Mills theory
(Seiberg-Witten ef\/fective theory) the following remarkable system of
generalized WDVV equations emer\-ged \cite{martini:1,martini:2}: 
\begin{equation} 
\label{martini:eq.1} F_i F^{-1}_k F_j = F_j F^{-1}_k F_i \qquad i,j,k = 1,
\ldots , n, 
\end{equation} 
where $F_i$ is the matrix
\[ 
(F_i)_{mr} = \frac{\partial^3 F}{\partial a_i \partial a_m \partial
a_r}
\] 
of third order derivatives of a function $F(a_1 , \ldots , a_n)$.
 
This system of nonlinear equations is satisf\/ied by the Seiberg-Witten
prepotential def\/ining the low-energy ef\/fective action. Moreover the
leading perturbative approximation to this exact Seiberg-Witten
prepotential should satisfy this set of equations by itself. For
instance for the gauge group $SU(n)$ the expression
\[ 
F_{\mbox{pert}} = \frac14 \sum_{i\leq i < j\leq n-1} (a_i - a_j)^2
\log(a_i - a_j)^2 + \frac12 \sum^{n-1}_{i=1} a^2_i \log a^2_i
\] 
def\/ines a solution of the generalized WDVV-system (\ref{martini:eq.1}).
 
Of course other gauge groups may be considered and more general
solutions may be proposed for classical Lie groups
\cite{martini:3,martini:4}. So although extremely 
dif\/f\/icult to solve in general, this overdetermined system of
nonlinear equations admit exact solutions. In this note we shall
present a complete proof that a substantial class of solutions for
the system (\ref{martini:eq.1}) may be constructed from root systems of
semisimple Lie algebras.

\section{Solutions from root systems}
Actually we have the following result.
\begin{theorem} 
\label{martini:T1} 
Let $R$ be the root system of a semisimple Lie algebra \textgoth{g}.
Then the function
\begin{equation} 
\label{martini:eq.2} 
F(a) = \frac14 \sum_{\alpha \in R} (\alpha, a)^2 \log (\alpha , a)^2
\end{equation} 
defined on the Cartan subalgebra \textgoth{h} of \textgoth{g}
satisfies the generalized WDVV equations~(\ref{martini:eq.1}). 
 
Here the bracket represents the Killing form of \textgoth{g}.
\end{theorem} 
 
In order to prove this theorem we show that we can rewrite the system
(\ref{martini:eq.1}) into an equivalent form which is more suitable for our
purposes.
 
\begin{proposition} 
Let $G=\sum\limits^n_{i=1} c_i F_i$ be an invertible linear
combination of the matrices $F_i$ with coefficients $c_i$ which may
depend on $a$.
 
Then $F$ is a solution of the WDVV-system (\ref{martini:eq.1}) if and only if
\begin{equation} 
\label{martini:eq.3} F_i G^{-1} F_j = F_j G^{-1} F_i \qquad i,j=1,\ldots , n.
\end{equation} 
\end{proposition} 
 
{\bf Proof.} Suppose $F$ satisf\/ies the WDVV-system (\ref{martini:eq.1}). Then
by inverting these equations we obtain
\[ 
F^{-1}_j F_k F^{-1}_i = F^{-1}_i F_k F^{-1}_j.
\] 
 
By taking linear combinations we get
\[ 
F^{-1}_j G F_i^{-1} = F^{-1}_i G F^{-1}_j.
\] 
 
Inverting once more yields the equations (\ref{martini:eq.3}). To prove the
converse we set $C_i = G^{-1}F_i$. Then (\ref{martini:eq.3}) implies that
$C_i$ and $C_j$ commute.  So 
\[ 
G^{-1} F_i F_k^{-1} F_j = C_i C^{-1}_k C_j = C_j C_k^{-1} C_i =
G^{-1} F_j F_k^{-1} F_i.
\] 
 
Thus $F$ is a solution of the WDVV-system (\ref{martini:eq.1}).
 
We continu by proving our main result, theorem \ref{martini:T1}. Without
restriction we may suppose that the root system is irreducible. Let
$\alpha_1, \ldots , \alpha_n$ be a basis of the Cartan subalgebra
\textgoth{h} of the Lie algebra \textgoth{g} consisting of simple
roots. Moreover let $a=\sum\limits^n_{i=1} a_i \alpha_i$.
 
For the linear combination $G=\sum^n_{i=1} a_i F_i$, where $F$ is
given in (\ref{martini:eq.2}), we have
\[ 
\hspace*{-5.2pt}
G_{km} = \sum_i a_i F_{ikm} = \sum_i a_i \sum_{\alpha \in R}
\frac{(\alpha, \alpha_i)(\alpha, \alpha_k)(\alpha,
\alpha_m)}{(\alpha, a)} 
= \sum_{\alpha \in R} (\alpha, \alpha_k)(\alpha , \alpha_m) =
(\alpha_k , \alpha_m)
\] 
using the very def\/inition of the Killing form. So in this case $G$
equals the matrix of the Killing form on a basis of simple roots.
 
For this choice of $G$ we have
\[ 
\hspace*{-5.2pt}
\left(F_i G^{-1} F_j\right)_{rs} = 4\sum_{\alpha , \beta \in R^+}
\frac{(\alpha, \alpha_i)(\alpha, \alpha_r)(\alpha, \beta)(\beta,
\alpha_j)(\beta , \alpha_s)}{(\alpha, a)(\beta, a)},
\] 
where $R^+$ denotes the positive part of the root system.
Consequently
\[ 
\hspace*{-8.6pt}
\left(F_i G^{-1} F_j - F_j G^{-1} F_i\right)_{rs} 
= 4\sum_{\alpha, \beta \in R^+}\frac{(\alpha, \beta)(\alpha,
\alpha_r)(\beta, \alpha_s)[(\alpha, \alpha_i)(\beta,
\alpha_j)-(\alpha, \alpha_j)(\beta, \alpha_i)]}{(\alpha, a)(\beta,
a)} 
\] 
 
We have to prove that this last expression vanishes, but by a close
inspection we see that it is antisymmetric in $r,s$. Therefore we may
also prove that
\[ 
\left(F_i G^{-1} F_j - F_j G^{-1} F_i\right)_{rs} 
- \left(F_i G^{-1} F_j - F_j G^{-1}F_i\right)_{sr}
\] 
vanishes. This last expression equals
\[ 
4\sum_{\alpha, \beta \in R^+} \frac{(\alpha, \beta)[(\alpha,
\alpha_i)(\beta, \alpha_j)-(\alpha, \alpha_j)(\beta,
\alpha_i)][(\alpha, \alpha_r)(\beta, \alpha_s)-(\alpha, 
\alpha_s)(\beta, \alpha_r)]}{(\alpha, a)(\beta, a)} 
\] 
which we abbreviate to
\begin{equation} 
\label{martini:eq.4} =\sum_{\{\alpha, \beta\}} t_{\{\alpha, \beta \}}, 
\end{equation} 
where $\{\alpha, \beta\}$ denotes an unordered pair of dif\/ferent
roots $\alpha, \beta$ in $R^+$. To f\/inish our proof we consider two
separate cases. First we consider the case that the Lie algebra
\textgoth{g} is simply-laced. In this case the roots have equal
length which we suppose to be normalized so that the squared 
lengths equal 2.
 
Now consider a (unordered) pair of roots $\{ \alpha, \beta\}$ in
$R^+$ such that $(\alpha, \beta )<0$. Then since the roots have equal
length it follows (see table 1, p.45 in \cite{martini:5}) that $(\alpha,
\beta)=-1$ and therefore (lemma 2, p.45 in \cite{martini:5}) that
$\alpha + \beta$ is again a root in $R^+$. Moreover: $(\alpha ,
\alpha+\beta)=1$, $(\beta, \alpha+\beta)=1$. Conversely if $\{
\alpha',\beta'\}$ is a pair of roots in $R^+$ such that $(\alpha',
\beta')>0$ then necessarily $(\alpha' , 
\beta')=1$ and a small calculation shows that there is a unique pair of roots $\{\alpha, \beta\}$ in $R^+$ such that $\{\alpha', \beta'\}=\{\alpha, \alpha+\beta\}$ or 
$\{\alpha' , \beta'\} = \{ \beta , \alpha+\beta\}$.
 
Consequently in this simply-laced case the sum
$\sum\limits_{\{\alpha, \beta\}} t_{\{\alpha, \beta\}}$ may be split
up into a sum of expressions of the form
\begin{equation} 
\label{martini:eq.5} t_{\{\alpha, \beta\}} + t_{\{\alpha, \alpha +\beta\}} +
t_{\{ \beta, \alpha + \beta\}},
\end{equation} 
where $\{ \alpha, \beta\}$ represents an anordered pair of roots in
$R^+$ with $(\alpha, \beta)=-1$. Using the relation
\[ 
\frac{1}{(\alpha, a)(\beta, a)} = \frac{1}{(\alpha, a)(\alpha+\beta,
a)} + \frac{1}{(\beta, a)(\alpha + \beta, a)} 
\] 
it is now easy to see that the expression (\ref{martini:eq.5}) and with it
the sum (\ref{martini:eq.4}) vanishes. This completes the proof of the
theorem in the simply-laced case.
 
In the non simply-laced case we have to consider also pair of roots
of unequal length.
 
First observe that when the root system is of type $G_2$ the
dimension of the Cartan subalgebra equals 2 and therefore the theorem
becomes trivial. So we may ignore this special case. In the other non
simply-laced cases the ratio of the squared length of a long and a
short root equals 2. We may assume the length of the short root to be
equal to 1.
 
Now consider a pair $\{\alpha, \beta\}$ of roots in $R^+$ with
$\alpha$ a short and $\beta$ a long root such that $(\alpha,
\beta)<0$. Then it follows (table 1, p.
45 in \cite{martini:5}) that $(\alpha, \beta) = -1$. We construct the
$\alpha$-string through $\beta$. It consists of $\beta , \beta +
\alpha, \beta + 2\alpha$. For the inner
product of the roots $\alpha, \beta, \beta + \alpha, \beta + 2\alpha$
we have
\[ 
(\beta , \beta + \alpha)=(\beta + \alpha, \beta + 2\alpha)=(\alpha,
\beta + 2\alpha) = 1
\] 
and
\[ 
(\beta , \beta + 2\alpha)=(\alpha, \beta + \alpha)=0.
\] 
 
We obtain three pairs $\{\alpha+\beta ,\beta\}, \{\alpha+\beta,
2\alpha + \beta\},\{\alpha, 2\alpha+\beta\}$ of a short and a long
root with inner product equal to 1. Conversely, a simple calculation
shows that each pair $\{ \alpha' , \beta'\}$ with $\alpha'$ short,
$\beta'$ long and $(\alpha', \beta')=1$ is obtained uniquely in this
way.Using the relation
\[ 
\frac{1}{(\alpha, a)(\beta, a)} =  
\frac{1}{(\alpha + \beta, a)(\beta , a)} + 
\frac{1}{(\alpha + \beta, a)(2\alpha + \beta , a)} + 
\frac{1}{(\alpha, a)(2\alpha + \beta , a)} 
\] 
it is now easily seen that
\[ 
t_{\{\alpha, \beta\}} + t_{\{\alpha+\beta , \beta\}} +
t_{\{\alpha+\beta , 2\alpha+\beta\}} + t_{\{\alpha,
2\alpha+\beta\}}=0.
\] 
 
For pairs of equal length we may argue as in the simply-laced case.
Consequently the sum $\sum\limits_{\{\alpha, \beta\}} t_{\{\alpha,
\beta\}}$ split up into
vanishing parts consisting of three or four terms. This completes the
proof of the theorem in the non simply-laced case.
 
As is well-known a root system is the weight system of the adjoint
representation of some semisimple Lie algebra \textgoth{g}. So one
may try to construct more general solutions in terms of weight
systems.
 
Actually generalizing the expression (\ref{martini:eq.2}) in a
natural way we obtain for any representation $\varphi$ of the Lie algebra
\textgoth{g} the function
\begin{equation} 
\label{martini:eq.6} F_{\varphi} (a) = \sum_{\lambda \in W}
\lambda(a)^2 \log \lambda(a)^2 
\end{equation} 
def\/ined on the Cartan subalgebra \textgoth{h} of \textgoth{g}. Here
summation is over the set of weights of the representation
$\varphi$.

From experience with concrete representations we know that
unfortunately in general formula (\ref{martini:eq.6}) does not satisfy the
WDVV equations despite the fact that for some representations the
WDVV equations indeed hold. For a short review from the point of view
of physics see e.g. \cite{martini:4}. To our best knowledge at present no
precise results are known in the literature. We hope to report on
these questions in the near future.
} 
 
 \label{martini-lp}
\end{document}